\documentclass[twocolumn,showpacs,floatfix,nofootinbib]{revtex4}

\usepackage{graphicx}
\usepackage{bm}
\usepackage{amsmath}
\usepackage{amssymb}

\newcommand{\vect}[1]{\bm{#1}}

\begin{document}
\title{Lifetime of the A($v'=0$) state and Franck-Condon factor of the A-X(0-0) transition of CaF measured by the saturation of laser-induced fluorescence}

\author{T. E. Wall$^{1}$}
\author{J. F. Kanem$^{1}$}
\author{J. J. Hudson$^{1}$}
\author{B. E. Sauer$^{1}$}
\author{D. Cho$^{2}$}
\author{M. G. Boshier$^{3}$}
\author{E. A. Hinds$^{1}$}
\author{M. R. Tarbutt$^{1}$}

\affiliation{$^{1}$Centre for Cold Matter, Blackett Laboratory, Imperial
College London, London SW7 2AZ, United Kingdom.\\
$^{2}$Department of Physics, Korea University, Seoul 136-713, Korea.\\
$^{3}$Physics Division, Los Alamos National Laboratory, MS D454, Los Alamos, New Mexico 87545, USA.}

\date{\today}

\begin{abstract}
We describe a method for determining the radiative decay properties of a molecule by studying the saturation of laser-induced fluorescence and the associated power broadening of spectral lines. The fluorescence saturates because the molecules decay to states that are not resonant with the laser. The amplitudes and widths of two hyperfine components of a spectral line are measured over a range of laser intensities and the results compared to a model of the laser-molecule interaction. Using this method we measure the lifetime of the A($v'=0$) state of CaF to be $\tau\!=\!19.2 \pm 0.7$\,ns, and the Franck-Condon factor for the transition to the X($v=0$) state to be $Z\!=\!0.987 {+0.013\atop -0.019}$. In addition, our analysis provides a measure of the hyperfine interval in the lowest-lying state of A($v'=0$), $\Delta_{e}\!=\!4.8 \pm 1.1$\,MHz.
\end{abstract}
\pacs{33.50.Dq, 33.70.Ca, 33.15.Pw}
%33.50.Dq 	Fluorescence and phosphorescence spectra
%33.70.Ca 	Oscillator and band strengths, lifetimes, transition moments, and Franck–Condon factors
%33.15.Pw 	Fine and hyperfine structure
\maketitle

\section{Introduction} \label{Sec:Introduction}

Atomic and molecular beams are often detected by laser-induced fluorescence (LIF). The technique offers high efficiency, excellent spectral resolution, and, if the beam is pulsed, high temporal resolution \cite{Tarbutt(1)08}. In passing through a probe laser whose frequency is resonant with an electronic transition, the molecules scatter photons from the laser into a detector. If the overall detection efficiency and the mean number of photons scattered per molecule are known, the absolute molecular flux can be determined. The mean number of photons scattered per molecule is a function of laser intensity, but saturates when this intensity is high enough. In this paper, we study this saturation of laser-induced fluorescence in detail, and we use our results to measure the natural linewidth and the Franck-Condon factor of the A-X(0-0) transition in CaF.

We distinguish two mechanisms for the saturation of laser-induced fluorescence. In the first mechanism, saturation occurs once the excitation rate $R$ exceeds the spontaneous decay rate $\Gamma$. Further increasing the intensity does not increase the fluorescence, since this becomes limited by the rate of spontaneous emission. This first mechanism is most important for `closed' systems where the excited state can only decay back to the initial state. In an `open' system, which is more common, the excited state can decay either to the initial state or to one or more other states that are not coupled to the laser. In this case, the average number of photons per molecule cannot exceed $1/(1-p)$, where $p$ is the probability of returning to the initial state. The first saturation mechanism still applies to this case, but now the fluorescence will also saturate when $R T \gg 1/(1-p)$, for then, in the time $T$ taken to pass through the probe laser, the average number of scattered photons per molecule has reached its maximum value. Further increasing $R$ by increasing the laser intensity cannot then increase the fluorescence. The first mechanism is treated in most text books on atom-laser interactions, e.g. \cite{Loudon}, but it is the second mechanism that is responsible for saturating laser-induced fluorescence in practically all atomic and molecular beam experiments that use cw lasers. The detection of molecular beams by laser-induced fluorescence is treated by Hefter and Bergmann \cite{HefterChapterScoles}, and in the earlier review articles by Kinsey \cite{Kinsey(1)77} and by Altkorn and Zare \cite{Altkorn(1)84}.

When fluorescence saturates by either mechanism, the LIF spectral lines will broaden because the degree of saturation will be greater on the line centre than in the wings. This power broadening is very well known in the context of the first saturation mechanism, though again it is the second mechanism that causes power broadening in most high resolution LIF spectroscopy with atomic and molecular beams. By measuring the power broadened linewidth as a function of probe intensity the natural linewidth of the transition, and hence the excited state lifetime, can be determined, provided that other broadening mechanisms are well controlled. It is also possible to determine the value of $p$ by measuring the saturation of fluorescence with probe intensity. For an electronic transition in a diatomic molecule, $p$ will depend on rotational factors that are usually straightforward to calculate, and on the Franck-Condon factor for the transition which is often unknown. Thus, by measuring how the widths of the spectral lines broaden and how the amplitudes of the lines saturate as the probe intensity is increased we can determine lifetimes and Franck-Condon factors. These are important molecular parameters that are often rather difficult to measure with high precision. Their determination has received renewed interest recently because of the possibility of laser cooling certain molecules where the lifetime is short and the Franck-Condon factor is very close to unity \cite{DiRosa(1)04}.

Lifetimes are usually measured directly by observing the rate of fluorescence decay following excitation by a short laser pulse. For CaF, the lifetime of the A($v'=0$) state was measured to be $21.9 \pm 4$\,ns by this method \cite{Dagdigian(1)74}. Franck-Condon factors are sometimes measured directly by measuring the relative intensities of a sequence of vibrational transitions, but more often are determined theoretically using potential energy curves constructed from measured spectroscopic parameters. In the case of CaF, both parameters have been calculated recently using the multireference configuration interaction method with a large basis set \cite{Pelegrini(1)05}, so our measurements provide a precise test of this advanced molecular theory. The technique we use here is a general one that could be applied to many other molecules.

\section{Experiment} \label{Sec:Experimental}

\begin{figure}
\includegraphics{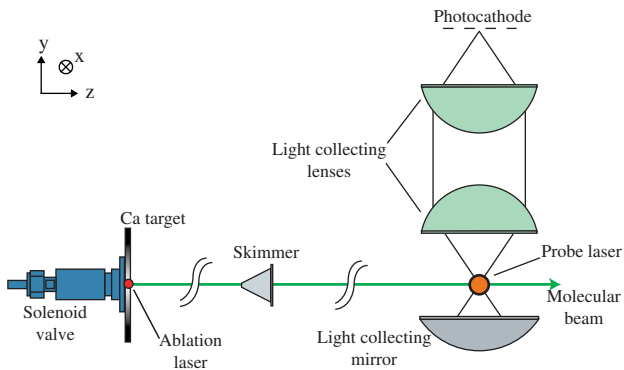}
\caption{
\label{Fig:Setup}
(Color online) Experiment setup}
\end{figure}

The experimental method is rather simple. The setup is similar to the one described in \cite{Tarbutt(1)02}, and is shown in Fig.\,\ref{Fig:Setup}. A pulsed solenoid valve filled with a mixture of Ar (98\%) and SF$_{6}$ (2\%) to a pressure of 4\,bar, emits gas pulses of 100-200\,$\mu$s duration at a rate of 10\,Hz into a vacuum chamber maintained below $10^{-4}$\,mbar. Immediately outside the nozzle of this valve, at the moment when the gas density is highest, a Ca target is ablated by light propagating along $x$ and produced by a Nd:YAG laser that provides pulses of duration 10\,ns, energy 25\,mJ and wavelength 1064\,nm. This laser pulse defines the zero of time in the experiment. The Ca target is made in an inert environment by cutting strips of Ca, $\sim 2$\,mm thick and $\sim 4$\,mm wide, and then gluing them to the rim of a stainless steel disk. This disk is mounted in the vacuum chamber with its axis parallel to $z$, and with the Ca surface displaced a few mm from the beamline in the $x$ direction. The flux of CaF decays, so after a few minutes of ablating a single spot of the target, it is rotated a little so that a fresh piece of Ca is exposed to the laser. The beam passes through a 2\,mm diameter skimmer, situated 70\,mm downstream, into a second vacuum chamber maintained below $10^{-7}$\,mbar, and then through a detector situated 810\,mm from the source.

In the detector, the time-of-flight profile of the CaF molecular pulse is recorded, with a resolution of 10\,$\mu$s, by laser-induced fluorescence at 606.3\,nm. Light from the interaction region is imaged, with a magnification of 2, onto the 28\,mm diameter photocathode of a photomultiplier tube operated in current mode. A rectangular mask, 9\,mm along $x$ and 20\,mm along $z$ is placed in front of the photocatode to spatially filter the fluorescence. This limits the visible part of the molecular beam to a 4.5\,mm region along $x$, thereby reducing the Doppler contribution to the measured linewidth to about 5\,MHz. This is important for an accurate determination of the natural linewidth. The probe laser beam comes from a tunable dye laser, passes first through an acousto-optic modulator (AOM), then through beam-shaping optics, then polarizing optics and finally propagates through the vacuum chamber (along $x$) via Brewster-angled windows. The AOM is used to modulate or scan the amplitude or the frequency of the light, the beam-shaping optics are used to produce a collimated circular beam approximately 2\,mm in diameter with an approximately top-hat intensity distribution, and the polarization optics ensure that the laser light is plane polarized along $z$. The power of the probe laser is monitored just before it enters the vacuum chamber, and the beam profile monitored by splitting off a portion of the beam onto a profiler whose distance from the beamsplitter is the same as that of the laser-molecule interaction region.

The time-of-flight profile is used to determine the molecular speed distribution whose mean is typically 600\,m/s. Integrating the time of flight profile gives the total fluorescence, which we measure as a function of laser frequency to obtain a spectrum. We analyze laser-induced fluorescence spectra that contain just two spectral lines, the $F\!=\!1$ and $F\!=\!0$ hyperfine components of the A-X(0--0)$Q_{11}(1/2)$ transition, separated by 123\,MHz. A portion of the laser beam passes into an evacuated confocal cavity whose free spectral range and finesse are 250\,MHz and 2.1 respectively. The transmission of this cavity is recorded at the same time as the spectra, and this signal is used to linearize the spectral data and calibrate the frequency axis.

\begin{figure}
\includegraphics{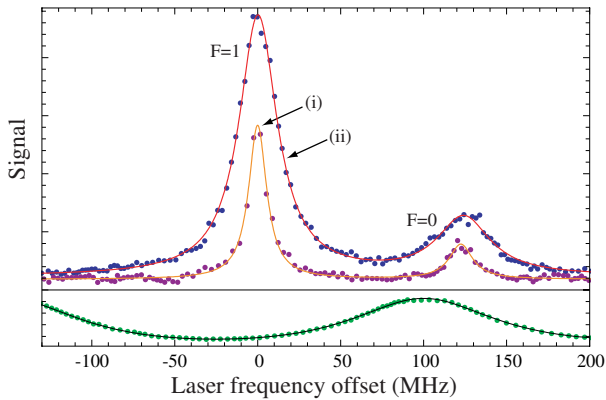}
\caption{
\label{Fig:Scan}
(Color online) Points, upper: typical laser-induced fluorescence spectra showing the two hyperfine components of the CaF A-X$(0\!-\!0)\,Q_{11}(\tfrac{1}{2})$ transition. The intensity at the centre of the laser beam was 11.5\,W/m$^{2}$ for spectrum (i) and 97.6\,W/m$^{2}$ for (ii). Lines, upper: double Lorentzian fits to the data. Points, lower: transmission of a 250\,MHz confocal cavity used to linearize the scan and calibrate the frequency axis. Line, lower: Airy function fit to the cavity data.}
\end{figure}

Figure \ref{Fig:Scan} shows example spectra obtained at two different probe intensities, along with typical cavity data. The line through the cavity data is a fit to an Airy function, and the line through each spectrum is a sum of two Lorentzians, with the amplitude, centre and width of each Lorentzian floating in the fit. Because the Doppler broadening is very small in the experiment, this is the expected lineshape for sufficiently low laser intensities. At higher laser intensities the spectral lines are power broadened and the exact lineshape can only be determined by the detailed modelling discussed in Sec.~\ref{Sec:Model}. Nevertheless, the two-Lorentzian lineshape was found empirically to be a good fit to all the data used in our analysis. At very high laser intensities the lineshape departs from the two-Lorentzian model. In Fig.~\ref{Fig:Scan}, the amplitude ratio between the $F\!=\!1$ and $F\!=\!0$ lines is not 3, as it would be from the degeneracy factor alone, but close to 4.5 for both spectra. This happens because an excited $F\!=\!1$ molecule is more likely to decay back to the original hyperfine state and scatter a second photon than is the case for an $F\!=\!0$ molecule. For spectrum (i), for which $I_{0}=11.5$\,W/m$^{2}$, the full width at half maximum (FWHM) is 17\,MHz for the $F\!=\!0$ line and 13\,MHz for the $F\!=\!1$ line. For the $I_{0}=97.6$\,W/m$^{2}$ data, the lines are strongly power broadened and the widths have increased to 39\,MHz and 28\,MHz. The power broadening is a direct result of the saturation in the scattered photon number, which in turn is due to the decay of excited molecules to states that are not resonant with the laser. Because $F\!=\!1$ molecules can scatter more laser photons, the onset of saturation occurs at higher intensity than for $F\!=\!0$ molecules, and so at a given intensity the $F\!=\!1$ linewidth is smaller.

Spectra such as those shown in Figure \ref{Fig:Scan} were analyzed over a wide range of laser intensities. For each spectrum, the scan was linearized, and then the amplitudes and widths of the two lines determined by fitting a two-Lorentzian model to the data. Fitting to Voigt functions instead of Lorentzians made a negligible difference to the fit residuals.

For the majority of the data used in our analysis, the AOM was used to modulate the probe power from high to low on successive shots from the source so that we obtained two (nearly) simultaneous spectra, one at high power and the other at a lower power. The high value was fixed for the entire set of spectral scans, while the lower value was varied between scans. This allowed us to normalize all the scans so as to eliminate slow drifts in the molecular flux. In other data sets we used the AOM to modulate the frequency of the probe laser by 123\,MHz while scanning the frequency of the dye laser as usual. This gave us two (nearly) simultaneous spectra, one shifted by 123\,MHz relative to the other, so that the $F\!=\!1$ to $F\!=\!0$ amplitude ratio could be measured with high precision. Finally, some additional data were obtained by locking the frequency of the dye laser and scanning the probe frequency using the AOM. This has the advantage that the frequency scan is exactly linear, without the need for later correction using the cavity data. Since the bandwidth of the AOM is limited, only one hyperfine component could be scanned this way, and then only for low laser intensities where the linewidth is small. For these data, the rf power delivered to the AOM was controlled by a feedback loop so as to keep the laser power constant over the scan.

\section{Model} \label{Sec:Model}

\begin{figure}
\includegraphics{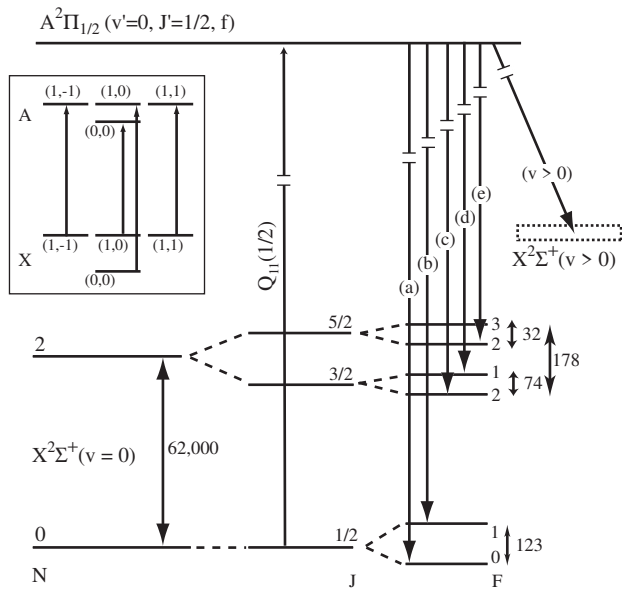}
\caption{
\label{Fig:EnergyLevels}
Relevant energy levels of CaF. Frequency intervals are shown in MHz and are taken from \cite{Childs(1)81} and \cite{Kaledin(1)99}. The plane-polarized laser drives the A--X(0--0)$Q_{11}(1/2)$ transition. The four magnetic-hyperfine components of the laser transition are shown in detail in the inset. The excited state can decay back to the $v\!=\!0$ ground state on any of the branches (a)--(e), with branching ratios given by Eq.~(\ref{Eq:branchingRatios}). Decay to other vibrational states of X are also possible and are indicated by $(v>0)$.}
\end{figure}

Figure \ref{Fig:EnergyLevels} shows the relevant molecular energy levels for the experiment. In the X\,$^{2}\Sigma^{+}\,(v\!=\!0)$ ground state, the angular momentum of the rotating nuclei, $\hat{\vect{N}}$, couples to the electron spin, $\hat{\vect{S}}$, to form a total electronic angular momentum $\hat{\vect{J}}=\hat{\vect{N}}+\hat{\vect{S}}$. Since $S=1/2$, each rotational level (apart from $N\!=\!0$) is split by the spin-rotation interaction into two components with $J\!=\!N\pm 1/2$. Coupling to the $I\!=\!1/2$ spin of the fluorine nucleus leads to further splitting of each level into hyperfine components labelled by $F$, the quantum number of the total angular momentum $\hat{\vect{F}}=\hat{\vect{I}}+\hat{\vect{J}}$. In the experiment, the laser drives $Q_{11}(1/2)$ transitions from the ground state, X\,$^{2}\Sigma^{+}\,(v\!=\!0,N\!=\!0,J\!=\!1/2)$, to the electronically excited state, labelled A\,$^{2}\Pi_{1/2}\,(v'\!=\!0,J'\!=\!1/2,f)$. Here, the orbital angular momentum, $\hat{\vect{L}}$, is strongly coupled to the internuclear axis, and so too is the electron spin due to a strong spin-orbit coupling. The sum of their projections, $\Omega=\Lambda+\Sigma$, has magnitude $1/2$, this being the spin-orbit component of lower energy. In the A state, the projection of $\hat{\vect{L}}$ onto the internuclear axis can either be $\Lambda'=+1$ or $\Lambda'=-1$. The energy eigenstates are the symmetric and antisymmetric superpositions of these two projections. The degeneracy between these two eigenstates, which have opposite parity, is lifted by a small degree of mixing with other electronic states ($\Lambda$-doubling). The laser drives transitions only to the state labelled $f$, the one of negative parity, because the ground state has positive parity. The hyperfine structure in the excited state is the same as in the ground state, but is much smaller and is not shown in the main part of the figure.

The excited state can decay to any of the vibrational states of X, but calculations indicate that decay back to the $v\!=\!0$ state is the most probable by far \cite{Pelegrini(1)05}. We use $Z$ to denote the Franck-Condon factor for this decay route. Within this channel, the decay may proceed either to the $N\!=\!0, J\!=\!1/2, F\!=\!0,1$ states, labelled (a) and (b) in the figure and known as the $Q_{11}(1/2)$ transition, or to the $N\!=2,J\!=\!3/2, F\!=\!2,1$ states, labelled (c) and (d) in the figure and known as the $P_{12}(3/2)$ transition. Because the hyperfine interaction is not diagonal in $J$, the two $F\!=\!2$ levels are of mixed $J$ character. It follows that decay to the upper $F\!=\!2$ level is also possible, and we label it (e). Note that decays to $N\!=\!1$ are forbidden since these states have negative parity, so we have not shown these levels in the figure. The decay to the $F\!=\!3$ level is forbidden by the $\Delta F = 0,\pm1$ selection rule. Transitions to all other vibrational states are labelled ($v>0$), and have branching ratio $1-Z$. The branching ratios for the decay channels (a)-(e) are calculated in Appendix \ref{AppA}.

Taking the quantization axis along the polarization direction ($z$), the laser drives only those transitions shown in the inset of Fig.~\ref{Fig:EnergyLevels}. We use the index $i$ to denote any of the four sublevels of the ground state, and the index $j$ to denote the corresponding sublevels in the excited state. The coherences reach a steady-state on the timescale of the excited-state lifetime. Since the interaction time is far longer than this, we can neglect the initial transients in the coherences. In this limit, the problem reduces to a set of rate equations with the laser excitation rate given by:
\begin{equation}
R(\delta)=\frac{\Gamma /2}{1+4 \delta ^2/\Gamma ^2}\frac{2\Omega ^2}{\Gamma ^2}=\frac{\Gamma /2}{1+4 \delta ^2/\Gamma ^2}\frac{I}{I_s}.
\label{Eq:R}
\end{equation}
Here, $\Gamma$ is the spontaneous decay rate of the upper level, $\delta=\omega_{L}-\omega_{ij}$ is the detuning of the laser angular frequency $\omega_{L}$ from the resonance angular frequency $\omega_{ij}$, $\Omega$ is the Rabi frequency, $I$ is the laser intensity, and $I_{s}$ is called the saturation intensity and is defined by the relation $I/I_s=2\Omega ^2/\Gamma ^2$. It is the characteristic intensity for the first saturation mechanism discussed in Sec.\,\ref{Sec:Introduction}.

The Rabi frequency for the transition is $\Omega=E_{0} z_{ij}/\hbar$, where $z_{ij}=\langle i |\hat{\vect{d}}.\hat{\vect{z}}|j\rangle$, the laser field is $\vect{E}=E_{0}\hat{\vect{z}}\cos(\omega_{L}t)$, $\hat{\vect{d}}$ is the electric dipole operator and $\hat{\vect{z}}$ is a unit vector along the $z$-axis. Using the relations $I=1/2 \epsilon _0 c E_0{}^2$, and
\begin{equation}
\Gamma =\frac{1}{\pi  \epsilon _0\hbar  c^3}\sum _k \omega _{jk}^{3}z_{jk} ^2,
\label{Eq:Gamma}
\end{equation}
we obtain
\begin{equation}
I_{s}=\frac{\pi  h c \Gamma }{\lambda ^3}\frac{\sum _k z_{jk}{}^2}{z_{ij}{}^2}.
\label{Eq:IsFinalForm}
\end{equation}
Here, the index $k$ denotes all possible final states in the decay of the excited state. We have used the fact that the radiation is isotropic (App.\,\ref{AppA}) and we have made the approximation that $\omega_{jk}=2\pi c/\lambda$, for all final states $k$. This is a good approximation for all relevant states within $v\!=\!0$ whose transition frequencies only differ by a part in $10^{4}$. It is less good for the $v>0$ decays whose frequencies differ by several percent, but since the values of $z_{jk}$ for these transitions is known to be very small \cite{Pelegrini(1)05}, the error contributed to $\Gamma$ by this approximation is negligible. We divide the states $k$ into those within the ground state $J\!=\!1/2$ manifold, labelled by $i$, and all other states. With this notation, we can write $r \sum _k z_{jk}{}^2 = \sum _i z_{ji}{}^2$, which defines the branching ratio, $r$, to the $J=1/2$ manifold. We now use the fact that, for each excited state sublevel labelled by $j$, there is only one non-zero matrix element in the sum over $i$, as illustrated in the inset of Fig.~\ref{Fig:EnergyLevels}. We thus obtain a convenient expression for the saturation intensity,
\begin{equation}
I_s=\frac{\pi  h c \Gamma }{\lambda ^3}\frac{3}{2Z},
\label{Eq:Isat}
\end{equation}
where we have taken the branching ratio $r=2Z/3$ from Eq.~(\ref{Eq:branchingRatios}).

We now write down the rate equations that govern how the populations in the various states shown in the inset of Fig.~\ref{Fig:EnergyLevels} evolve with time. The laser intensity is sufficiently low, and its frequency sufficiently far detuned from all the other states, that their populations do not need to be included in the rate equations. There is a symmetry between the $(1,-1)$ and $(1,1)$ states so we do not need separate rate equations for each. We denote the populations in the $(0,0)$ and $(1,0)$ ground states by $N_{\text{g0}}$ and $N_{\text{g1}}$, the sum of the populations in the $(1,1)$ and $(1,-1)$ ground states by $N_{\text{g2}}$, and the ground state hyperfine interval by $\Delta_{g}$. The notation for the excited state populations and hyperfine interval is analogous. With the laser detuning, $\delta$, referenced to the $(F'\!=\!1,M_{F}'\!=\!0) \leftarrow (F\!=\!0,M_{F}\!=\!0)$ transition frequency, the coupled rate equations for a molecule of speed $v$ are
\begin{align}
v\frac{dN_{\text{g0}}}{dz}&=R(\delta)[N_{\text{e1}}\!-\!N_{\text{g0}}]+\frac{r}{3} \Gamma[N_{\text{e1}}\!+\!N_{\text{e2}}],\label{Eq:RateEq1}\\
v\frac{dN_{\text{g1}}}{dz}&=R(\delta\!+\!\Delta_g\!+\!\Delta_e)[N_{\text{e0}}\!-\!N_{\text{g1}}]+\frac{r}{3}\Gamma[ N_{\text{e0}}\!+\!N_{\text{e2}}],\label{Eq:RateEq2}\\
v\frac{dN_{\text{g2}}}{dz}&=R(\delta\!+\!\Delta_g)[N_{\text{e2}}\!-\!N_{\text{g2}}]+
\frac{r}{3}\Gamma[2(N_{\text{e0}}\!+\!N_{\text{e1}})\!+\!N_{\text{e2}}],\label{Eq:RateEq3}\\
v\frac{dN_{\text{e0}}}{dz}&=R(\delta\!+\!\Delta_g\!+\!\Delta_e)[N_{\text{g1}}\!-\!N_{\text{e0}}]-\Gamma N_{\text{e0}},\label{Eq:RateEq4}\\
v\frac{dN_{\text{e1}}}{dz}&=R(\delta)[N_{\text{g0}}\!-\!N_{\text{e1}}]-\Gamma  N_{\text{e1}},\label{Eq:RateEq5}\\
v\frac{dN_{\text{e2}}}{dz}&=R(\delta\!+\!\Delta_g)[N_{\text{g2}}\!-\!N_{\text{e2}}]-\Gamma N_{\text{e2}}\label{Eq:RateEq6}.
\end{align}
Normalizing all populations to the total population initially in the $J\!=\!1/2$ ground state manifold, the initial conditions at the source are $N_{\text{g0}}=N_{\text{g1}}=\tfrac{1}{2}N_{\text{g2}}=\tfrac{1}{4}$ and $N_{\text{e0}}=N_{\text{e1}}=N_{\text{e2}}=0$.

Our model does not include the effect of the background magnetic field in the experiment. Using a fluxgate magnetometer this field was measured to be $B_{x}=-45 \pm 5$\,mG, $B_{y}=-212 \pm 2$\,mG and $B_{z}=-143 \pm 1$\,mG. The resulting Zeeman splitting between the $M_{F}\!=\!\pm 1$ sublevels is smaller than 1\,MHz in both ground and excited states. In the ground state, this is too small relative to the hyperfine interval for there to be any significant mixing of the two hyperfine components. It is also too small relative to the natural linewidth to have any influence on the excitation rates. The spin precession of the excited $F'\!=\!1$ state is too slow compared to the excited state lifetime to have any influence. The excited state hyperfine interval, $\Delta_{e}$, is not yet determined. If it is comparable to $\Gamma$, it is much larger than the Zeeman splitting, and so the mixing of the $F\!=\!1,0$ states is small; if it is much smaller than $\Gamma$ the time evolution of the excited state is again too slow to have any influence. These considerations show that our neglect of the magnetic field is justified.

The solution of the rate equations depends on the three molecular parameters to be determined, the decay rate $\Gamma$, the Franck-Condon factor $Z$, and the excited state hyperfine interval $\Delta_{e}$. Note from Eq.~(\ref{Eq:Isat}) that $I_{s}$ is determined once $\Gamma$ and $Z$ are known, and so does not constitute an extra unknown parameter. To solve the rate equations, we also need to know the $z$-dependence of $R$, which enters via the intensity profile of the laser. This was carefully monitored during the experiments. For the model, we describe the laser profile by the function
\begin{equation}
I(z,y)=I_{0}\frac{\tanh(r/d\!+\!L/d)-\tanh(r/d\!-\!L/d)}{2\tanh(L/d)},
\end{equation}
where $r=\sqrt{z^{2}+\epsilon^{2}y^{2}}$, $L$ and $\epsilon$ describe the size and roundness of the beam, and $d$ describes the slope of the profile at the edges. This function looks like a top hat when $d \ll L$ and like a Gaussian when $d\simeq L$, and we use it to fit all of our measured beam profiles. They typically have $L\simeq 1.5$\,mm, $d\simeq 0.4$\,mm and $\epsilon\simeq 1$.

\begin{figure}
\includegraphics{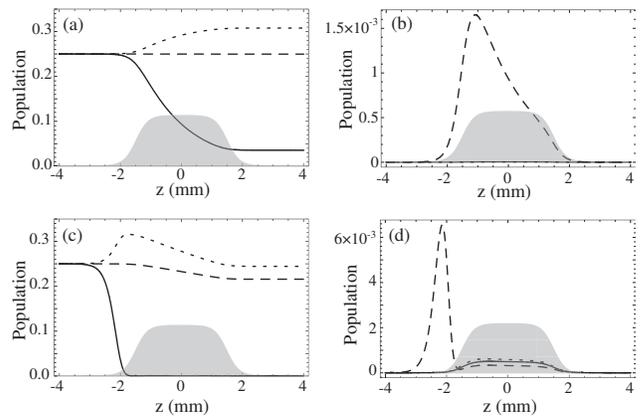}
\caption{
\label{Fig:Populations}
Ground and excited state populations for a molecule traveling along the beam axis at 600\,m/s and passing through a laser beam resonant with the transition from $F=0$. Solid lines show the $(0,0)$ populations, dashed lines the $(1,0)$ populations, and dotted lines the $(1,\pm1)$ populations. The grey area indicates the profile of the laser beam. The parameters of the laser are $L=1.5$\,mm, $d=0.4$\,mm and $\epsilon=1$, and the molecular parameters are chosen as $\Gamma=2\pi\times 8$\,MHz, $Z=0.96$ and $\Delta_{e}=0$. (a) and (b) show, respectively, the ground and excited state populations when $I_{0}=0.02I_{s}$. In (c) and (d) they are shown for $I_{0}=4I_{s}$.}
\end{figure}

Figure \ref{Fig:Populations} shows how the populations change with $z$ as a molecule passes through the laser beam centred at $z=0$ and resonant with the transition from $F=0$. In (a) and (b) the laser is weak, $I_{0}=0.02I_{s}$. The ground state $(0,0)$ population falls as the molecule passes through the beam (solid line in (a)), and some small population appears in the $(1,0)$ component of the excited state (dashed line in (b)). Some of this excited state population decays to the $(1,\pm1)$  ground states, and so their population rises (dotted line in (a)). The laser is too weak to excite the $F=1$ states, whose transition frequencies are detuned by many natural linewidths. In (c) and (d) the laser is much stronger, $I_{0}=4I_{s}$, and so the population in the resonant $(0,0)$ state falls to zero as soon as the molecule starts to enter the beam. Again, this excitation is reflected in the population of the $(1,0)$ excited state (dashed line in (d)), and some of this decays to the $(1,\pm1)$ ground states as before, resulting in an initial increase in their populations. In addition, the laser is now intense enough that it can excite population out of the $F=1$ levels, even though they are far detuned. Consequently, their populations also fall, though at a much slower rate than that of the $F=0$ level. This off-resonant excitation is reflected in the excited state populations shown in (d) which all become non-zero and follow the profile of the laser. The excited state population is greatest in $(1,\pm1)$ (dotted line), simply because the corresponding ground states have more population. The small pedestal in the $(1,0)$ excited state population occurs because population excited first to $(1,\pm1)$ can decay to the $(0,0)$ ground state and then be rapidly re-excited to $(1,0)$ by the resonant laser.

The number of photons scattered by a molecule is found by integrating the excited state populations over time, and multiplying by $\Gamma$. As shown in App.~\ref{AppA}, photons are emitted isotropically in the spontaneous decay of all four excited states so the signal at the detector is directly proportional to the scattered photon number without any further complication. In the example discussed above, the average number of photons scattered per ground-state molecule is $0.28$ when $I=0.02I_{s}$ and $0.82$ when $I=4I_{s}$. By contrast, when the laser is instead tuned to the $F=1$ resonance, these values change to $1.07$ and $1.89$ respectively. To verify the accuracy of our rate equation approach and the neglect of the magnetic field, we reproduced the results shown in Fig. \ref{Fig:Populations} using a full density matrix calculation that includes all the relevant coherences in the problem. The results obtained by the two methods were in excellent agreement. The full calculation reduces the scattered photon number by 0.4\% in the low intensity case, and increases it by 0.1\% in the high intensity case. These differences have a negligible effect on our final results.

To model the experiment fully, we first account for the divergence of the molecular beam in both the horizontal and vertical planes. A molecule that arrives at the laser at position $(x,y)$ will see the laser frequency Doppler-shifted by $\delta\omega_{\text{dopp}}=-2\pi v x/(\lambda D)$, and will see less power than a molecule at $y=0$ because of the laser profile. Molecules with $|x| > 2.25$\,mm do not contribute to the signal since they are outside the detection region, while molecules displaced too far in the vertical direction see hardly any power and so also contribute negligibly to the signal. We solve the rate equations to determine the fluorescence signal for many different values of $x$ and $y$ within these boundaries, and average the results to obtain the final signal at the detector. Repetition of this procedure for many different values of $\delta$ results in a model spectrum. To this we fit a double Lorentzian so that it is analyzed in exactly the same manner as the corresponding experimental spectrum. A set of model spectra obtained for a range of laser intensities constitutes a complete simulation of the experiment. Comparison of the experimental data to the simulation data for various $\Gamma$, $Z$ and $\Delta_{e}$, determines these three quantities.

\section{Analysis} \label{Sec:Results}

\begin{figure}
\includegraphics{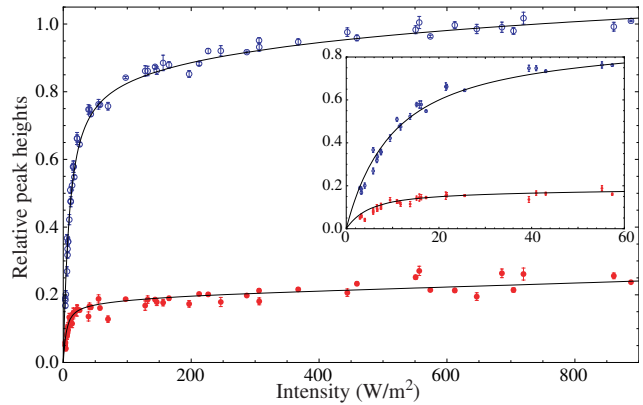}
\caption{
\label{Fig:Amplitudes}
(Color online) Peak heights of $F\!=\!0$ and $F\!=\!1$ lines determined by fitting two Lorentzians, as a function of $I_{0}$, the laser intensity at the centre of the probe. The other parameters of the probe beam are $d=0.38$\,mm, $L=1.18$\,mm, $\epsilon = 1$. Filled circles: experimental data for $F\!=\!0$. Open circles: experimental data for $F\!=\!1$. Solid lines: corresponding results of the model, using the best-fit values $Z=0.987$, $\Gamma/(2\pi)=8.29$\,MHz, $\Delta_{e}=4.8$\,MHz. The inset shows the low intensity region in greater detail. Data from several experimental runs have been combined.}
\end{figure}

Figure \ref{Fig:Amplitudes} shows the measured peak heights of the $F\!=\!1$ and $F\!=\!0$ spectral lines as a function of laser intensity, together with the corresponding results of the model. As the intensity is increased from zero, there is initially a very rapid increase in the amplitudes of both lines. In this regime, the excitation rate, $R$, is low, so that $R T \le 1$, where $T$ is the transit time of molecules through the probe beam. At higher laser intensities, the rapid increase of the amplitudes ceases because all the resonant molecules have decayed to other states not resonant with the laser. As seen most clearly in the inset, where only the low intensity region is plotted, this saturation occurs at lower intensity for the $F\!=\!0$ molecules because, after excitation, they are less likely to return to a resonant state than is the case for $F\!=\!1$. Inspection of Eq.~(\ref{Eq:branchingRatios}) shows that the probabilities are $2Z/9$ for $(0,0)$, $4Z/9$ for $(1,\pm 1)$ and $6Z/9$ for $(1,0)$. For this same reason, the amplitude ratio reaches a maximum value of about $4.5$, as already discussed in the context of Fig.~\ref{Fig:Scan}. It is also clear from Fig.~\ref{Fig:Amplitudes} that the peak heights have not completely saturated at high intensity, but instead continue to increase with laser intensity, though with a far smaller gradient. This is due to the off-resonant excitation of the other hyperfine component, for which the excitation rate is small over the entire intensity range shown in the plot. This has a greater influence on the $F\!=\!0$ line, there being more $F\!=\!1$ molecules to excite off-resonance, so the fractional rate of increase is larger for this line. At very high intensities, approximately 20 times higher than the maximum intensity shown in the plot, the amplitudes stop increasing and complete saturation has occurred. At this point, molecules in both hyperfine components scatter photons until they decay elsewhere, irrespective of which hyperfine component is resonant with the probe. Then, the mean number of photons scattered per ground state molecule would be $1/(1-r)=3$ if $Z$ were exactly 1.

Figure \ref{Fig:Amplitudes} shows that the model reproduces the amplitude data very well over the entire range of intensities. In fitting the model to the data, the values of $Z$, $\Gamma$ and $\Delta_{e}$ are free parameters to be determined. We explain later how we determine these. There is also one additional free parameter, an overall scaling of the peak heights, which is not determined in the experiment since we do not measure the absolute flux of molecules. Both the $F\!=\!0$ and $F\!=\!1$ peak heights are scaled by the same value so as to find the best fit to the model. No scaling has been applied to the intensity axis. It is worth noting that all the details of the model we have outlined need to be included in order to obtain good agreement with the experimental data. For example, if all molecules are assumed to see the same laser intensity, rather than averaging over the known intensity profile, the `corner' where the on-resonance excitation saturates is far sharper in the model than we observe. It is also essential to generate complete synthetic spectra, and then fit to these, rather than simply determining the amplitudes at the two resonance frequencies.

\begin{figure}
\includegraphics{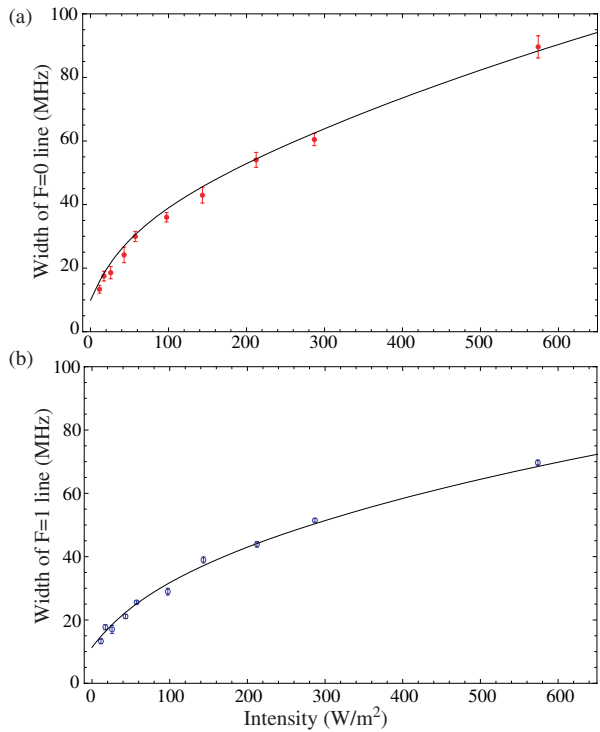}
\caption{
\label{Fig:Widths}
(Color online) Widths (FWHM), determined from two-Lorentzian fits, of (a) $F\!=\!0$ and (b) $F\!=\!1$ lines, as a function of $I_{0}$, the laser intensity at the centre of the probe. The other parameters of the probe beam are $d=0.38$\,mm, $L=1.18$\,mm, $\epsilon = 1$. The solid lines show the results of the model, using the best-fit values $Z=0.987$, $\Gamma/(2\pi)=8.29$\,MHz, $\Delta_{e}=4.8$\,MHz.}
\end{figure}

Figure \ref{Fig:Widths} shows how the widths of the two lines depend on the probe intensity. The observed power broadening is related to the saturation of the laser induced fluorescence discussed above: as the intensity is increased, the wings of the lines increase in amplitude more rapidly than the centres, and the lines broaden. It is immediately evident in the figure that the $F\!=\!0$ widths are larger than those of $F\!=\!1$, and that this difference increases with increasing intensity. This is because the onset of saturation occurs at lower intensity for $F\!=\!0$ as discussed above. The figure shows that the model again agrees very well with the data for the particular values of $Z$, $\Gamma$ and $\Delta_{e}$ used here. There are no additional free parameters; in particular no scaling is applied to either ordinate or abscissa. We have only shown the data obtained from one experimental run in the figure. Importantly, the widths obtained by scanning the AOM frequency (not shown) are consistent with the data shown here, indicating that our procedure for linearizing the scans is accurate.

We turn now to the evaluation of the molecular parameters. The dependence of the peak heights on intensity is primarily sensitive to the value of $Z$, as we would expect since this determines the mean number of photons scattered per molecule. Sensitivity of the amplitudes to $\Gamma$ arises via $I_{s}$ (see Eq.~(\ref{Eq:IsFinalForm})), and through the degree of off-resonance excitation at high intensity. However, given the freedom to scale the amplitudes, the sensitivity to $\Gamma$ is rather weak. The widths, on the other hand, are determined primarily by $\Gamma$ and are only weakly sensitive to $Z$. The excited state hyperfine interval, $\Delta_{e}$, is unresolved in all our spectra, but it does broaden the $F\!=\!1$ line while leaving the $F\!=\!0$ linewidth unchanged (see the inset to Fig.~\ref{Fig:EnergyLevels}). Furthermore, a larger value of $\Delta_{e}$ slightly decreases the gradient of the $F\!=\!1$ saturation curve at low intensity, while leaving the corresponding $F\!=\!0$ curve unaltered. This happens because the laser ceases to be exactly resonant with all three sub-components of $F\!=\!1$ simultaneously. Thus both the width data and the peak height data have a small degree of sensitivity to $\Delta_{e}$. To determine the central values of all three parameters, we simultaneously fit the amplitudes and widths of both lines to the model, weighting each data point by the inverse square of its standard error, and so find the parameters that minimize $\chi^{2}$ in this simultaneous fit. This procedure gives $Z\!=\!0.987 \pm 0.012$, $\Gamma/(2\pi)\!=\!8.29 \pm 0.08$\,MHz and $\Delta_{e}\!=\!4.8 \pm 0.4$\,MHz, where we have, so far, quoted only the statistical uncertainties. In addition to this simultaneous fit, we have also analyzed the amplitudes and widths separately. Fitting to just the amplitudes gives $Z=0.98$ and lacks sensitivity to the other two parameters as discussed above. Fitting to just the $F'\!=\!0$ widths gives $\Gamma/(2\pi)=8.0$\,MHz, whereas the fit to the $F'\!=\!1$ widths gives $\Gamma/(2\pi)=8.5$\,MHz.

To complete the analysis, we consider the systematic uncertainties. There is a 5\% uncertainty in the calibration of the power meter used to measure the laser power. Systematically shifting the intensity values, we find that this translates to $\pm 0.2$\,MHz in $\Gamma/(2\pi)$ but does not significantly alter the other two parameters. The zero-offset in the power measurements, and its uncertainty, are negligible. Drifts in the molecular flux can result in a systematic error in the ratio of the two resonant peak heights. We used the frequency-modulated data to estimate the size of the systematic drift of beam intensity between measurements of the two hyperfine components, and find it to be consistent with zero and less than 1.5\%. This upper bound to the drift contributes uncertainties of $+0.013\atop -0.015$ to $Z$, $\pm 1$\,MHz to $\Delta_{e}$ and a negligible amount to $\Gamma$. Another important uncertainty is the contribution of Doppler broadening to the measured linewidths arising from the divergence of the molecular beam. The Doppler broadening is accounted for in our determination of $\Gamma$ since our model integrates over the molecular beam divergence. In the experiment, an aperture in front of the photocathode defines the visible portion of the beam, and hence the beam divergence. An aperture placed in the molecular beam would have been better. We rely on the imaging property of the detection optics to define an effective aperture of $4.5$\,mm in the $x$-direction. To this, we assign a 20\% uncertainty. Modelling this, we find that decreasing the effective aperture by 20\% slightly decreases the linewidths over the entire intensity range, but since the Doppler broadening is smaller than the natural linewidth the best-fit value of $\Gamma/(2\pi)$ is only reduced by 0.2\,MHz. Increasing the effective aperture by 20\% increases the linewidths at low intensity as expected, but decreases them at high intensity. This is because, on average, more power is needed to reach a given level of saturation when the divergence increases, and so the degree of power broadening is reduced. The overall effect is again to reduce the best-fit value of $\Gamma/(2\pi)$ by 0.2\,MHz. A 10\% systematic uncertainty in the diameter of the probe beam leads to $\pm 0.14$\,MHz uncertainty in $\Gamma/(2\pi)$ and a negligible error in the other two parameters. Errors introduced in linearizing and calibrating the frequency axis can be divided into random errors, which are already included in the error bars on the individual data points, and a systematic error. Based on the consistency with the data obtained by scanning the AOM frequency and the accuracy we obtain for the known ground-state hyperfine interval, we estimate that the latter is negligible.

Combining the error estimates, we obtain the following final results: $Z\!=\!0.987 {+0.013\atop -0.019}$, $\Gamma/(2\pi)\!=\!8.29 \pm 0.30$\,MHz and $\Delta_{e}\!=\!4.8 \pm 1.1$\,MHz. From $\Gamma$, we find the excited state lifetime to be $\tau\!=\!19.2 \pm 0.7$\,ns. Our results are consistent with the theoretical values, $Z\!=\!0.964$ and $\tau\!=\!19.48$\,ns, given in \cite{Pelegrini(1)05}. Our measurement of the lifetime also agrees with the previous measurement \cite{Dagdigian(1)74}, $21.9 \pm 4$\,ns, and is 6 times more precise.

\section{Conclusions}

By studying in detail the power broadening and saturation of fluorescence on the A-X(0-0)\,Q$_{11}(1/2)$ transition in CaF, we have measured the excited state lifetime, the Franck-Condon factor, and the hyperfine interval in the excited state. Our results confirm the accuracy of the theoretical work of \cite{Pelegrini(1)05}. Our technique is rather general and can be used to measure the same parameters in other molecules. To determine the molecular parameters accurately, the data needed to be compared with synthetic spectra generated using a detailed rate model of the experiment. The model needs to include all the relevant states, and to take into account the laser beam profile and the Doppler shifts of off-axis molecules. Using our central values for $Z$ and $\Gamma$ in Eq.~(\ref{Eq:IsFinalForm}) we find the saturation intensity relevant to the first saturation mechanism, $I_{s}=222$\,W/m$^{2}$. Referring back to Fig.~\ref{Fig:Amplitudes} we see that (apart from the off-resonant excitation) the onset of saturation occurs at much lower intensity than this, confirming that it is indeed the second mechanism that is primarily responsible for saturating the fluorescence.

Our model can be used as the starting point for modelling more elaborate experiments that use multiple laser frequencies to access several states simultaneously, in order to scatter more photons. In this context it is interesting to speculate on the prospects of laser cooling a molecule with a structure like that of CaF. Figure \ref{Fig:EnergyLevels} shows that 5 laser frequencies (transitions (a)-(e)) are required to close off all the transitions within the $v=0$ manifold. For CaF, this would require two separate lasers to bridge the 62\,GHz gap between $N=0$ and $N=2$, and three AOMs to bridge the hyperfine intervals. Note that the system may pump into a dark state unless this is destabilized by applying a suitable magnetic field or modulating the laser polarization \cite{Berkeland(1)02}. Note also that, as more `leaks' are shut off, the laser intensity required to saturate the transition increases in proportion to the scattered photon number, until the intensity reaches $I_{s}$ at which point the first saturation mechanism takes over. Only under these conditions will the molecules scatter the maximum possible number of photons, which for our central value of $Z=0.987$ is 77 photons. If this could be achieved, each molecule could be detected with near unit efficiency, since the combined efficiency of PMT and collection optics is typically 2\%. This photon number is still very far from the number required for efficient laser cooling - for that, a value of $Z$ much closer to unity is required, or further laser frequencies are needed to close off the remaining leaks to $v>0$ \cite{DiRosa(1)04}. Speculating a little further, we note that polar molecules such as CaF can be brought close to rest using a Stark decelerator \cite{Bethlem(1)99, Tarbutt(1)04} and then confined in a trap for several seconds \cite{Bethlem(1)00}. In that case, molecules will already be at temperatures in the 10-100\,mK range and only a few hundred photon scattering events are needed to reach ultracold temperatures.

\acknowledgements
We are grateful to Tim Steimle for his help and advice with this paper. We thank Jon Dyne for his expert technical assistance. The work was supported by the EPSRC and by the Royal Society.

\appendix
\section{Branching ratios and angular distribution of spontaneous emission}
\label{AppA}
We calculate the branching ratios for the various decay channels shown in figure \ref{Fig:EnergyLevels}. The calculation is done in a basis of states labelled $|\Lambda,S,\Sigma,\Omega,J,I,F,M_{F}\rangle$, usually called the $(a_{\beta})$ basis and defined in Sec.~\ref{Sec:Model} in the context of the A state. The excited state is the one of negative parity. Omitting the quantum numbers beyond $J$, this state is
\begin{equation}
|f\rangle = \tfrac{1}{\sqrt{2}}|1,\tfrac{1}{2},-\tfrac{1}{2},\tfrac{1}{2},\tfrac{1}{2}\rangle - \tfrac{1}{\sqrt{2}}|-1,\tfrac{1}{2},\tfrac{1}{2},-\tfrac{1}{2},\tfrac{1}{2}\rangle. \nonumber
\end{equation}
In $X$, the expansion of the two states of interest in the case $(a_{\beta})$ basis is found from Eq.\,(6.149) of \cite{BrownCarrington}:
\begin{align}
|N=0,J=\tfrac{1}{2}\rangle &= \tfrac{1}{\sqrt{2}}|0,\tfrac{1}{2},\tfrac{1}{2},\tfrac{1}{2},\tfrac{1}{2}\rangle + \tfrac{1}{\sqrt{2}}|0,\tfrac{1}{2},-\tfrac{1}{2},-\tfrac{1}{2},\tfrac{1}{2}\rangle,\nonumber \\
|N=2,J=\tfrac{3}{2}\rangle &= \tfrac{1}{\sqrt{2}}|0,\tfrac{1}{2},\tfrac{1}{2},\tfrac{1}{2},\tfrac{3}{2}\rangle -\tfrac{1}{\sqrt{2}}|0,\tfrac{1}{2},-\tfrac{1}{2},-\tfrac{1}{2},\tfrac{3}{2}\rangle.\nonumber
\end{align}

Next we deal with the coupling of the electronic angular momentum and the nuclear spin. All the relevant states are diagonal in $J$ apart from the two states that have $F=2$, as discussed in Sec.~\ref{Sec:Model}. Diagonalization of this part of the problem can be done exclusively within the $N=2$ manifold following the procedure given in the appendix of \cite{Sauer(1)96} and using the spin-rotation and hyperfine parameters given in \cite{Childs(1)81}. Labelling the upper and lower $F=2$ levels by $|2\pm\rangle$, the result is\footnote{That the value of the coefficient, numerically 0.4995, should be so close to $\tfrac{1}{2}$ seems to be coincidental.}

\begin{align}
|2+\rangle = \tfrac{1}{2}|J=\tfrac{3}{2},I=\tfrac{1}{2},F=2\rangle + \tfrac{\sqrt{3}}{2}|J=\tfrac{5}{2},I=\tfrac{1}{2},F=2\rangle,\nonumber\\
|2-\rangle = \tfrac{\sqrt{3}}{2}|J=\tfrac{3}{2},I=\tfrac{1}{2},F=2\rangle - \tfrac{1}{2}|J=\tfrac{5}{2},I=\tfrac{1}{2},F=2\rangle.\nonumber
\end{align}

Now we calculate the matrix elements of the electric dipole operator, written in spherical tensor notation as $T^{(1)}(\vect{d})$, between basis states of $A$ and $X$. Applying the Wigner-Eckart theorem and noting that the operator acts on the electronic angular momentum, the matrix elements factorize to
\begin{equation}
m_{fi}=\langle f|T^{(1)}_{p}(\vect{d})|i \rangle = m_{1}\,m_{2}\,m_{3}
\end{equation}
where $p=0,\pm1$ labels the polarization, and
\begin{align}
m_{1}&=(-1)^{F'-M_{F}'}\left(
\begin{array}{ccc}
 F' & 1 & F \\
 -M_{F}' & p & M_{F}
\end{array}
\right),\\
m_{2}&=(-1)^{F+J'+I+1}\sqrt{(2F'+1)(2F+1)}\left\{
\begin{array}{ccc}
 J & F & I \\
 F' & J' & 1
\end{array}
\right\},\\
m_{3}&=\langle \Lambda',S,\Sigma',\Omega',J'||T^{(1)}(\vect{d})||\Lambda,S,\Sigma,\Omega,J\rangle.
\end{align}
This can be further reduced by expressing the operator in the molecule-fixed coordinate system following  \cite{BrownCarrington}, in particular section 5.5.5, to reach
\begin{multline}
m_{3}=\sum _{q=-1}^1 (-1)^{J'-\Omega' }\sqrt{(2J'+1)(2J+1)}\left(
\begin{array}{ccc}
 J' & 1 & J \\
 -\Omega'  & q & \Omega
\end{array}
\right)\times\\
\left\langle \Lambda' ,S,\Sigma' |T_q{}^{(1)}(\vect{d})|\Lambda,S,\Sigma \right\rangle
\end{multline}

To obtain the branching ratios, we do not need to evaluate the final factor in the above equation since it is common to all the branches. Note that the electric dipole operator cannot change either the spin or its projection, so the only non-zero matrix elements in the above sum are those that have $\Sigma=\Sigma'$. Since $\Lambda=0$ whereas $\Lambda' = \pm 1$, it follows that $\Omega=\Omega' \pm 1$ and so the $q=0$ term in the sum is zero. Writing the transition matrix elements in the form $m_{3}(\Omega',\Omega)$, and inserting the values for all the quantum numbers fixed in the problem $(|\Lambda'|=1,\Lambda=0,|\Sigma'|=S=\tfrac{1}{2},J'=\tfrac{1}{2})$, we have

\begin{align}
&m_{3}(\tfrac{1}{2},\tfrac{1}{2})=m_{3}(-\tfrac{1}{2},-\tfrac{1}{2})=0,\\
&m_{3}(\tfrac{1}{2},-\tfrac{1}{2})=\sqrt{2(2J+1)}\left(
\begin{array}{ccc}
 \tfrac{1}{2} & 1 & J \\
 -\tfrac{1}{2} & 1 & -\tfrac{1}{2}
\end{array}
\right)m_{4}\\
&m_{3}(-\tfrac{1}{2},\tfrac{1}{2})=(-1)^{\tfrac{1}{2}-J}m_{3}(\tfrac{1}{2},-\tfrac{1}{2})\\
&m_4=\left\langle 1,\tfrac{1}{2},-\tfrac{1}{2}\left|T_{+1}{}^{(1)}(\vect{d})\right|0,\tfrac{1}{2},-\tfrac{1}{2}\right\rangle
\end{align}

Putting all this together, we get the branching ratios for the entire set of decay channels. For reference, the numerical results are:
\renewcommand\arraystretch{1.2}
\begin{equation}
\left(
\begin{array}{r|cccc}
   & \text{(0,0)} & \text{(1,-1)} & \text{(1,0)} & \text{(1,1)} \\ \hline
 \text{(0,0,0)} & 0 & \frac{2}{9} & \frac{2}{9} & \frac{2}{9} \\
 \text{(0,1,-1)} & \frac{2}{9} & \frac{2}{9} & \frac{2}{9} & 0 \\
 \text{(0,1,0)} & \frac{2}{9} & \frac{2}{9} & 0 & \frac{2}{9} \\
 \text{(0,1,1)} & \frac{2}{9} & 0 & \frac{2}{9} & \frac{2}{9} \\
 \text{(2,2-,-2)} & 0 & \frac{1}{8} & 0 & 0 \\
 \text{(2,2-,-1)} & 0 & \frac{1}{16} & \frac{1}{16} & 0 \\
 \text{(2,2-,0)} & 0 & \frac{1}{48} & \frac{1}{12} & \frac{1}{48} \\
 \text{(2,2-,1)} & 0 & 0 & \frac{1}{16} & \frac{1}{16} \\
 \text{(2,2-,2)} & 0 & 0 & 0 & \frac{1}{8} \\
 \text{(2,1,-1)} & \frac{1}{9} & \frac{1}{36} & \frac{1}{36} & 0 \\
 \text{(2,1,0)} & \frac{1}{9} & \frac{1}{36} & 0 & \frac{1}{36} \\
 \text{(2,1,1)} & \frac{1}{9} & 0 & \frac{1}{36} & \frac{1}{36} \\
 \text{(2,2+,-2)} & 0 & \frac{1}{24} & 0 & 0 \\
 \text{(2,2+,-1)} & 0 & \frac{1}{48} & \frac{1}{48} & 0 \\
 \text{(2,2+,0)} & 0 & \frac{1}{144} & \frac{1}{36} & \frac{1}{144} \\
 \text{(2,2+,1)} & 0 & 0 & \frac{1}{48} & \frac{1}{48} \\
 \text{(2,2+,2)} & 0 & 0 & 0 & \frac{1}{24}
\end{array}
\right).
\label{Eq:branchingRatios}
\end{equation}
Here, columns are labelled by the $(F',M_{F}')$ values in the upper state, and rows labelled by the $(N,F,M_{F})$ values in the lower state.

The angular distribution of the spontaneously emitted photons is proportional to $\sin^{2}\theta$ for $\Delta M_{F}\!=\!0$ transitions and to $(1\!+\!\cos^{2}\theta)/2$ for $\Delta M_{F}\!=\!\pm 1$ transitions, where $\theta$ measures the angle between the observation direction and the $z$-axis. For the decay of the $(1,0)$ and $(0,0)$ states, Eq.~(\ref{Eq:branchingRatios}) shows that the total branching ratios for $\Delta M_{F}\!=\!0,\pm1$ are each $1/3$ and so the emission is isotropic for these states. For the $(1,1)$ state the branching ratios are $1/2,1/3,1/6$ for $\Delta M_{F}\!=\!-1,0,+1$ transitions and so the emission from this state is also isotropic. For the $(1,-1)$ state these branching ratios are reversed and the emission is again isotropic.

\end{document}